\documentclass[final,1p,times]{elsarticle}

\usepackage{amsmath,amssymb}

\journal{Nuclear Physics A}

\begin{document}

\begin{frontmatter}



\title{Effects of the $\Lambda$(1405) on the Structure of Multi-Antikaonic Nuclei}


\author{Takumi Muto$^a$, Toshiki Maruyama$^b$, Toshitaka Tatsumi$^c$}
\address[label1]{Department of Physics, Chiba Institute of Technology,
2-1-1 Shibazono,Narashino, Chiba 275-0023, Japan}

\address[label2]{Advanced Science Research Center, Japan Atomic Energy Agency, Tokai, Ibaraki 319-1195, Japan}

\address[label3]{Department of Physics, Kyoto University, 
Kyoto 606-8502, Japan}

\begin{abstract}
The effects of the $\Lambda$(1405) ($\Lambda^\ast$) on the structure of the multi-antikaonic nucleus (MKN), in which several $K^-$ mesons are embedded to form deeply bound states, are considered based on chiral symmetry combined with a relativistic mean-field theory.  
It is shown that additional attraction resulting from the $\Lambda^\ast$ pole has a sizable contribution to not only the density profiles for the nucleons and $K^-$ mesons but also the ground state energy of the $K^-$ mesons and binding energy of the MKN as the number of the embedded $K^-$  mesons increases.
\end{abstract}


\end{frontmatter}


\section{Introduction}
\label{sec:intro}

Exploring multi-strangeness systems is an important aspect of understanding hadron dynamics in dense matter. Kaon condensation in neutron stars may exist as a strangeness-nonconserving system, where kaon condensates are spontaneously produced from normal matter through weak processes, $N+n\rightarrow N+p+K^-$, $N+e^-\rightarrow N+K^-+\nu_e$ ($N=p,n$)\cite{fmmt96}. Recently multi-antikaonic nuclei (abbreviated as MKN), where several antikaons ($K^-$ mesons) are bound in the ground state of the nucleus, have been investigated\cite{mmt09,gfgm08}, stimulated by the proposal to explore deeply bound kaonic nuclear states and subsequent theoretical and experimental studies\cite{gh08}. The MKN is a strangeness-conserving system and should be formed by embedding a $K^-$ meson in the nucleus through strong processes. 
Both the kaon-condensed state in neutron-star matter
and the MKN formed in experiments are cold, dense objects originating from the common $\bar K-N$ and $\bar K-\bar K$ interactions in dense matter, so that they may be closely related with each other. 

We have considered properties of the MKN within the framework of a relativistic mean-field theory (RMF) coupled with the nonlinear effective chiral Lagrangian\cite{mmt09}. 
It has been shown that the lowest $K^-$ energy, $\omega_{K^-}$,  increases as the number of embedded $K^-$ mesons, $|S|$, becomes large and that it enters into the subthreshold resonance region of the $\Lambda$(1405) ($\Lambda^\ast$), where $\omega_{K^-}\simeq m_{\Lambda^\ast}-m_N$ = 467 MeV. This is because the contribution to the energy from the repulsive $\bar K-\bar K$ interaction becomes sizable with the increase in $|S|$ as compared with the attractive $\bar K-N$ interaction. In this paper, we take into account the $\Lambda^\ast$-pole contribution as well as range terms  and study these effects on the structure of the MKN. 
 
\section{Formulation}
\label{sec:form}

A spherical symmetry is assumed for the MKN, and the mass number $A$, the number of protons $Z$, and the number $|S|$ of embedded $K^-$ mesons with the lowest energy $\omega_{K^-}$ are kept fixed.
We start with the effective chiral Lagrangian, which incorporates $s$-wave interactions between the (nonlinear) $\bar K$ mesons and nucleons of the scalar type simulated by the $KN$ sigma term, $\Sigma_{KN}$, and of the vector type (Tomozawa-Weinberg term).
 The nonlinear $K^-$ field $\Sigma$ is given as $\Sigma = \exp[2i(K^+T_{4+i5}+K^-T_{4-i5})/f]$, where $T_{4\pm i5} (\equiv T_4\pm iT_5)$ is the SU(3) generator and $f$ (= 93 MeV) the meson decay constant.  The $K^-$ field is represented as $K^-(r)=f\theta(r)/\sqrt{2}$ with $\theta(r)$ being the chiral angle in the condensate approximation\cite{mmt09}. 
These $\bar K-N$ interactions are replaced by those generated by the $\sigma$ and $\omega$, $\rho$ mesons-exchanges, respectively, within the RMF\cite{mmt09}. 

The thermodynamic potential $\Omega$ for the MKN is derived under a local density approximation for the nucleons\cite{mmt09}. 
The correction to the energy density, $\Delta \epsilon(r)$, from the $\Lambda^\ast$ is introduced through the second-order perturbation with respect to the axial current of hadrons, $\hat A_5^\mu=f\partial^\mu K^- + \cdots +(g_{\Lambda^\ast}/2)(\bar\Lambda^\ast\gamma^\mu p +{\rm h.c.})+\cdots$ with $g_{\Lambda^\ast}$ being the coupling constant for $K^- p \Lambda^\ast$ vertex: 
\begin{eqnarray}
\Delta \epsilon&=& -i \int d^4 z \langle x| T\widetilde\omega_{K^-}
\hat A_5^0(z)\widetilde\omega_{K^-}\hat A_5^0(0) |x\rangle \times \left(-\frac{1}{2}\sin^2\theta\right) \cr
&\overset{\textrm{real part}}{\Rightarrow} &
- \frac{1}{2}f^2\widetilde\omega_{K^-}^2\sin^2\theta\Bigg\lbrack\rho_p^s\Bigg\lbrace d_p+\frac{g_{\Lambda^\ast}^2}{2f^2}\frac{m_{\Lambda^\ast}-m_N-\omega_{K^-}}{(m_{\Lambda^\ast}-m_N-\omega_{K^-})^2+\gamma_{\Lambda^\ast}^2}\Bigg\rbrace+d_n\rho_n^s\Bigg\rbrack \ , 
\label{eq:dele2}
\end{eqnarray}
where the smooth parts $\propto d_p\rho_p^s, d_n\rho_n^s$ are the range terms with $\rho_p^s (r)$ ($\rho_n^s (r)$) being the scalar density of the proton (neutron) and the pole contribution comes from the $\Lambda^\ast$ with $\gamma_{\Lambda^\ast}$ being the width. These terms are absorbed into the effective nucleon masses. We call these contributions to the energy the second-order effects (SOE)\cite{fmmt96}. In Eq.~(\ref{eq:dele2}), $\widetilde\omega_{K^-}(r)$ [$\equiv\omega_{K^-}-V_{\rm Coul}(r)$] is the lowest energy of the $K^-$ shifted in the presence of the Coulomb potential. 
The parameters, $d_p$, $d_n$, $g_{\Lambda^\ast}$, and $\gamma_{\Lambda^\ast}$ are determined so as to reproduce the on-shell $s$-wave $K-N$ scattering lengths\cite{m81}.
The classical $K^-$ field equation is given from $\delta\Omega/\delta\theta=0$ as
\begin{eqnarray}
\nabla^2\theta(r)&=&\sin\theta(r)\Bigg\lbrack m_K^{\ast 2}(r)-2
\widetilde\omega_{K^-}(r)X_0(r) -\widetilde\omega_{K^-}^2(r)\cos\theta(r) \cr &-&\widetilde\omega_{K^-}^2(r)\cos\theta(r)\Bigg\lbrace\rho_p^s(r)\left( d_p+\frac{g_{\Lambda^\ast}^2}{2f^2}\frac{m_{\Lambda^\ast}-m_N-\omega_{K^-}}{(m_{\Lambda^\ast}-m_N-\omega_{K^-})^2+\gamma_{\Lambda^\ast}^2}\right)+d_n\rho_n^s(r)\Bigg\rbrace\Bigg\rbrack \ ,
\label{eq:theta}
\end{eqnarray}
where $m_K^{\ast 2}(r)$ (= $m_K^2-2g_{\sigma K}m_K\sigma(r)$) is the square of the effective mass of the $K^-$, and $X_0(r)$ (= $g_{\omega K}\omega_0(r)+g_{\rho K}R_0(r)$) represents the $\bar K-N$ vector interaction. In these quantities, $g_{iK}$ ($i=\sigma, \omega, \rho$) are the coupling constants, while $\sigma(r)$, $\omega_0(r)$, and $R_0(r)$ are the mean fields of the $\sigma$ meson and the time components of the $\omega$ and $\rho$ mesons, respectively.
Together with Eq.~(\ref{eq:theta}) one obtains the coupled equations of motion (EOM) for the other mesons $\sigma$, $\omega$, $\rho$, and the Poisson equation for the Coulomb potential $V_{\rm Coul}(r)$: 
\begin{subequations}\label{eq:eom}
\begin{eqnarray}
 -\nabla^2\sigma (r)+m_\sigma^2\sigma (r)&=&-\frac{dU}{d\sigma}(r)+g_{\sigma N}(\rho_p^s(r)+\rho_n^s(r))+2g_{\sigma K}m_Kf^2(1-\cos\theta(r)) \ ,  \label{eq:eom1}\\
 -\nabla^2\omega_0(r)+m_\omega^2\omega_0(r)&=&g_{\omega N}(\rho_p(r)+\rho_n (r))-2g_{\omega K}\widetilde\omega_{K^-}(r) f^2(1-\cos\theta(r)) \ ,  \label{eq:eom2}\\
 -\nabla^2 R_0(r)+m_\rho^2 R_0(r)&=&g_{\rho N}(\rho_p(r)-\rho_n(r))-2g_{\rho K} \widetilde\omega_{K^-}(r) f^2(1-\cos\theta(r)) \ , \label{eq:eom3}\\
\nabla^2 V_{\rm Coul}(r)&=&4\pi e^2(\rho_p(r)-\rho_{K^-}(r)) \ , 
 \label{eq:eom4}
\end{eqnarray}
\end{subequations}
where $\rho_i(r)$ ($i=p,n,K^-$) are the number densities and $g_{iN}$ ($i=\sigma, \omega, \rho$) the coupling constants. 
The coupled equations (\ref{eq:theta}) and (\ref{eq:eom1})$-$(\ref{eq:eom4}) are solved self-consistently, and the density distributions $\rho_i(r)$ and other quantities are obtained as functions of the radial distance $r$. 
\vspace{-0.2cm}

\section{Numerical Results}
\label{sec:result}
We take the $^{15}_8$O ($A$=15, $Z$=8) as a reference nucleus. The $K^-$ optical potential depth $U_K$ is chosen to be $U_K$ = $-$ 80 MeV.  
\vspace{-0.2cm}

\subsection{Density profiles} 
\label{subsec:density}
\begin{figure}[h]
\noindent
\begin{minipage}[l]{0.50\textwidth}
\begin{center}
\includegraphics[height=.28\textheight]{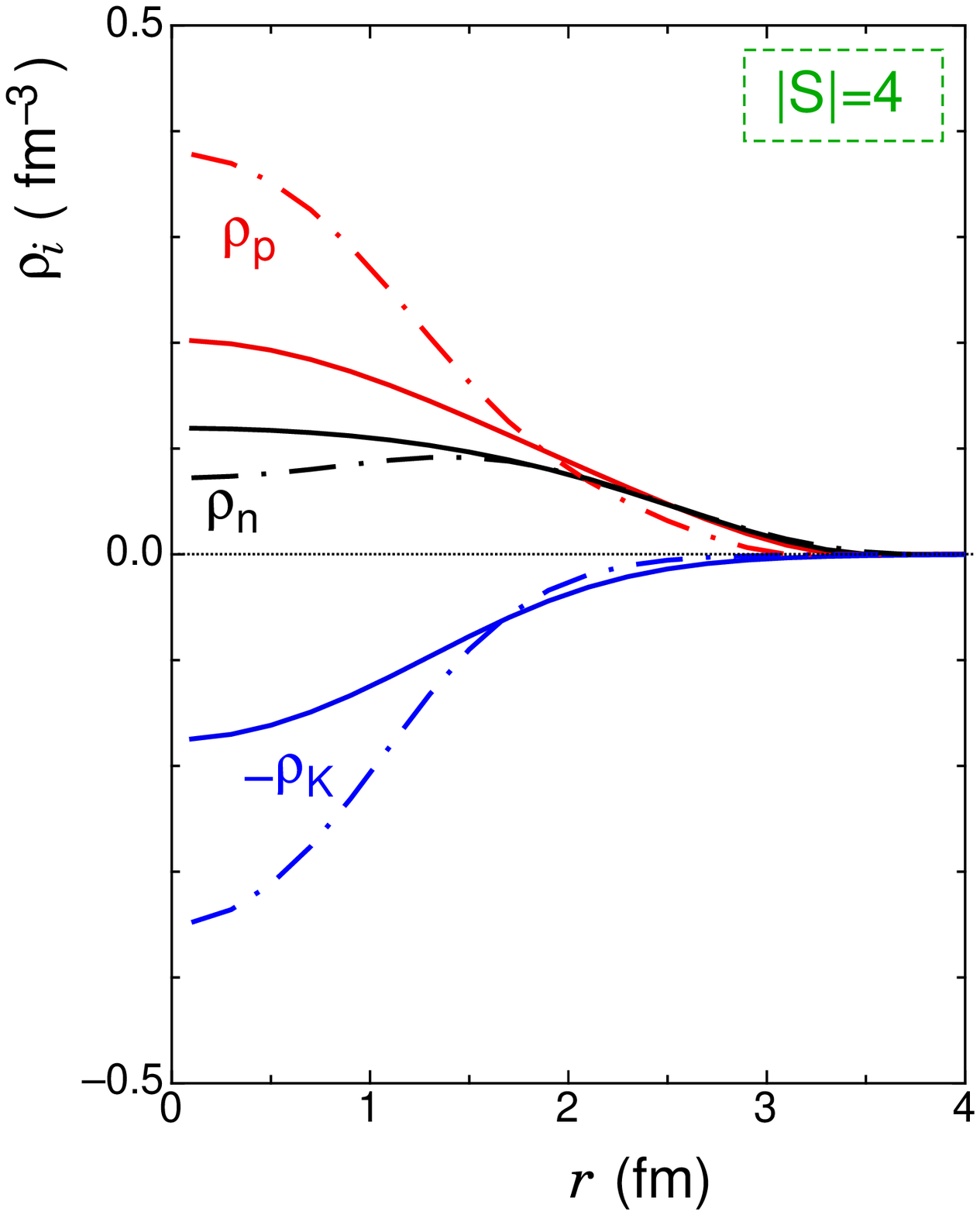}
\end{center}
\end{minipage}~
\begin{minipage}[r]{0.50\textwidth}
\begin{center}
\includegraphics[height=.28\textheight]{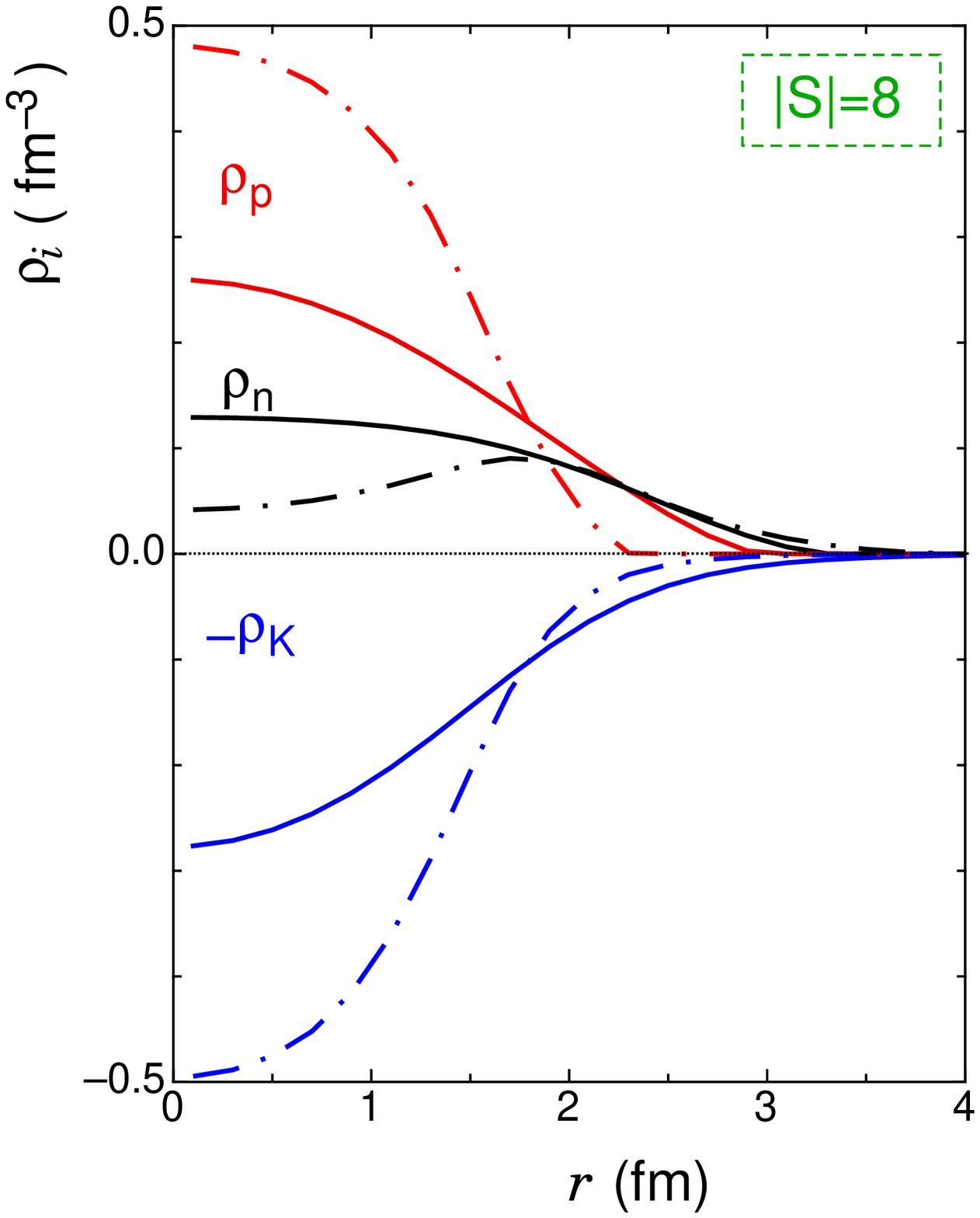}
\end{center}
\end{minipage}
\caption{The density distributions of protons, neutrons, and the  strangeness density [=$-\rho_{K^-} (r)$] for the MKN with $A$=15, $Z$=8, and $|S|$=4, 8 in the case of $U_K$=$-$80 MeV.  }
\label{fig1}
\end{figure}

The density distributions of the protons, neutrons, and the distribution of the strangeness density [=$-\rho_{K^-} (r)$] are shown for $|S|$=4 and 8 in Fig.~\ref{fig1}. The solid lines are for the previous result without the SOE\cite{mmt09}, and the dashed-dotted lines for the present result with the SOE.
Due to the SOE, the $K^-$ mesons and the protons are attracted more to each other than the case without the SOE, since in the former the $K^-$ lies below the resonance region of the $\Lambda^\ast$ and feels an additional attraction through coupling with the $\Lambda^\ast$ pole. As a result, the central densities of the protons and $K^-$ mesons become larger.  On the other hand, neutrons are pushed outward from the center of the MKN due to the weakly repulsive effect from the range term ($\propto d_n\rho_n^s$,\  $d_n<0$ in Eq.~(\ref{eq:theta}) ). These features become remarkable for a large value of $|S|$ (Compare the cases of $|S|$=4 and 8). The central baryon density $\rho_{\rm B}^{(0)}$ (=$\rho_p(r=0)+\rho_n(r=0)$) becomes $\rho_{\rm B}^{(0)}\sim$3.5 $\rho_0$ with $\rho_0$ = 0.153 fm$^{-3}$ for $|S|\sim$ 8. One can see a ``neutron skin'' structure with a thickness (1$-$2) fm for $|S|\sim$8. In addition, for a larger $|S|$, the proton and $K^-$ density distributions tend to be more uniform near the center. 
\vspace{-0.5cm}

\subsection{$|S|$-dependence of the lowest $K^-$ energy and binding energy }
\label{subsec:s}

\begin{figure}[h]
\noindent
\begin{minipage}[l]{0.50\textwidth}
\begin{center}
\includegraphics[height=.28\textheight]{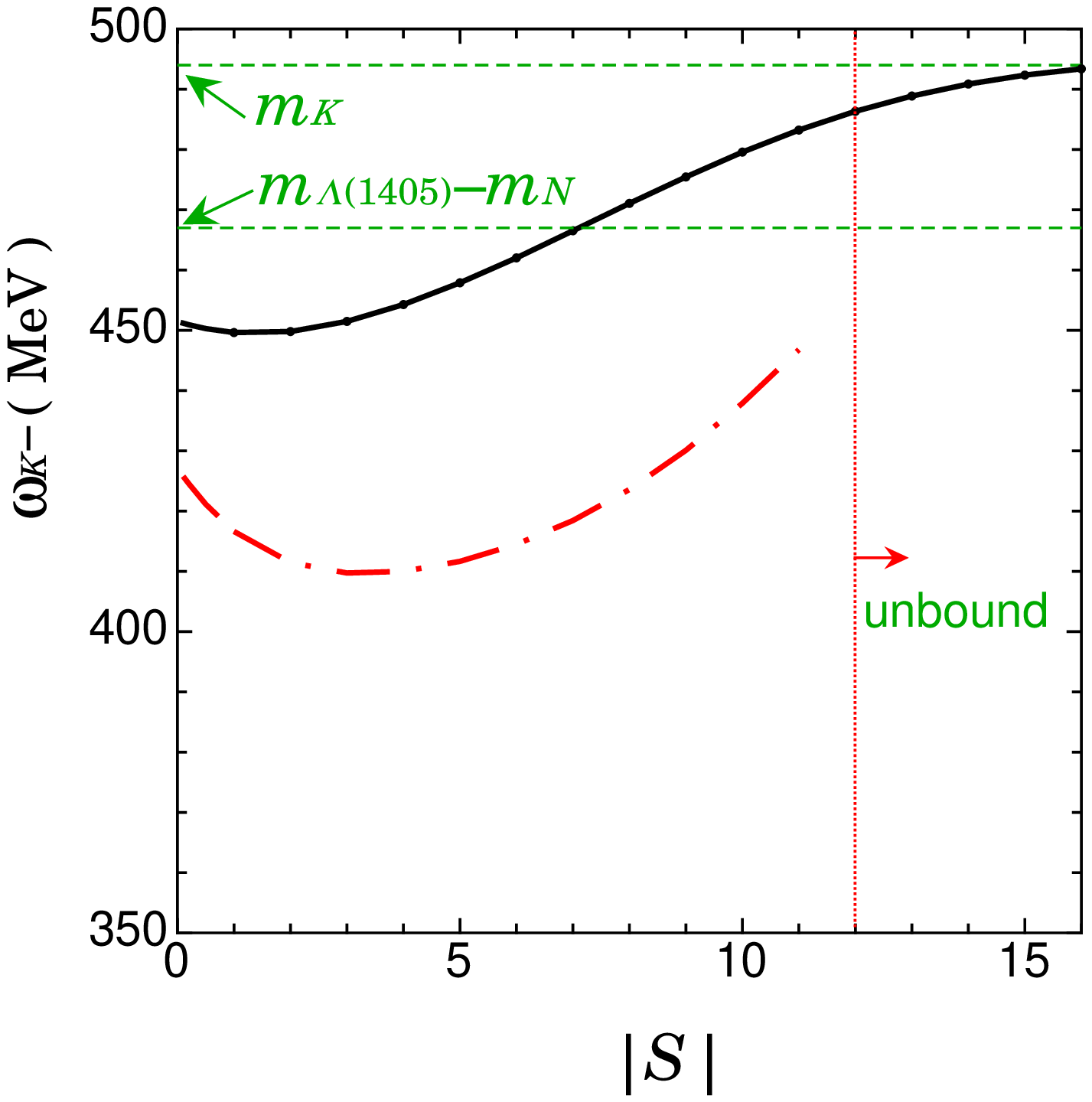}
\caption{The lowest energy of the $K^-$, $\omega_{K^-}$, for the MKN with $A$=15, $Z$=8, and $|S|$=2, 8 in the case of $U_K$=$-$80 MeV.  }
\label{fig2}
\end{center}
\end{minipage}~
\begin{minipage}[r]{0.50\textwidth}
\begin{center}
\includegraphics[height=.28\textheight]{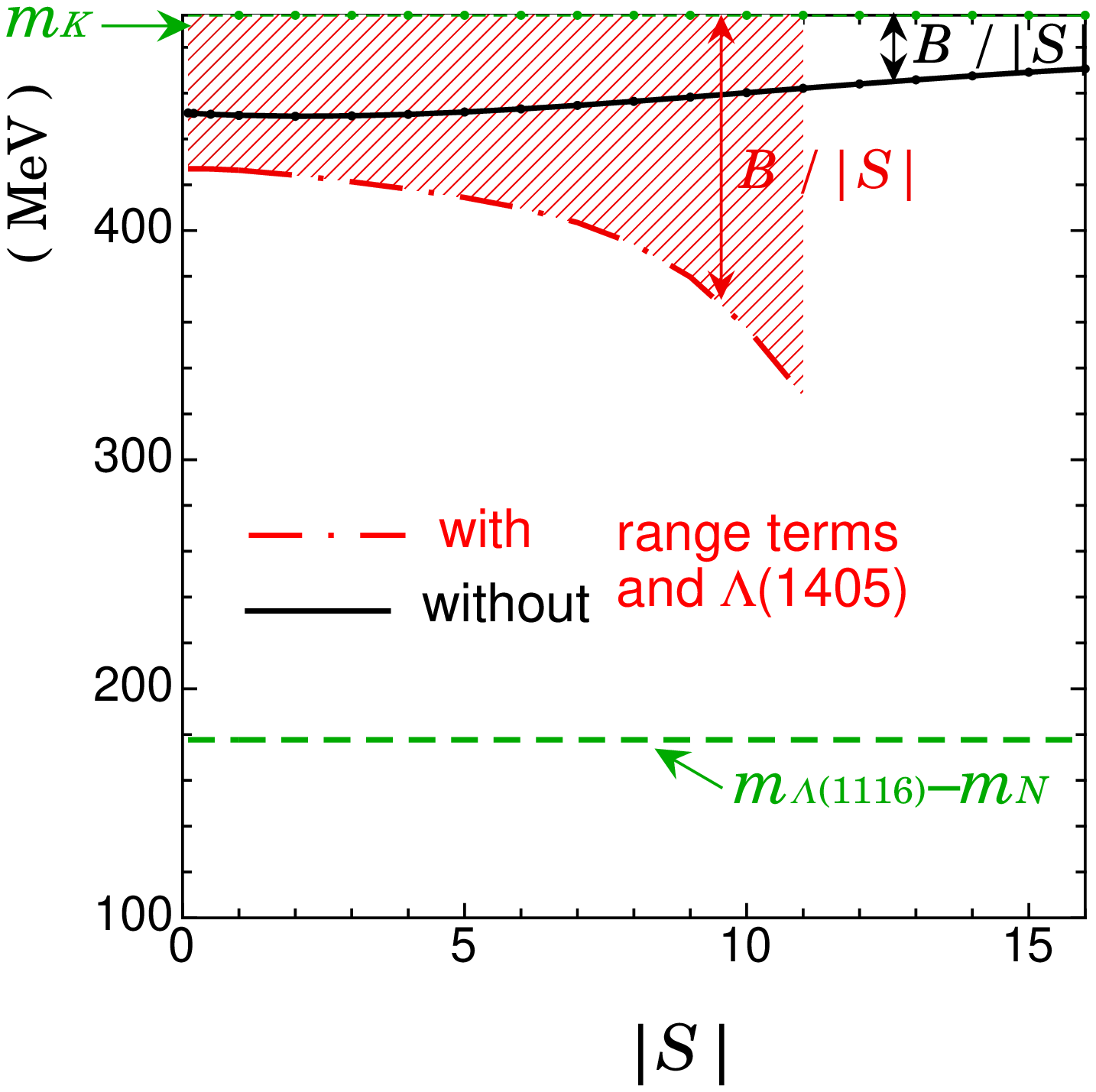}
\caption{The energy difference per $|S|$, $[E(A, Z, |S|)-E(A, Z, 0)]/|S|$ as functions of $|S|$.  $B(A, Z, |S|)$ is the binding energy of the MKN. }
\label{fig3}
\end{center}
\end{minipage}
\end{figure}

In Fig.~\ref{fig2}, the lowest energy of the $K^-$, $\omega_{K^-}$, is shown as a function of $|S|$. The energy difference per unit of strangeness, $[E(A, Z, |S|)-E(A, Z, 0)]/|S|$ (=$m_K-B(A, Z, |S|)/|S|$ with $B(A, Z, |S|)$ being the binding energy of the MKN), is shown as a function of $|S|$ in Fig.~\ref{fig3}. In these figures the solid lines are for the result without the SOE\cite{mmt09}, and the dashed-dotted lines for the result with the SOE.
From Fig.~\ref{fig2}, the $\omega_{K^-}$ is shown to be lowered by $\sim$ 40 MeV from that without the SOE due to the additional attraction brought about from the $\Lambda^\ast$ pole. Nevertheless, $\omega_{K^-}$ increases with an increase in $|S|$ since the repulsive $\bar K-\bar K$ interaction overwhelms the attractive $\bar K-N$ interactions at large $|S|$. 
 For $|S|\geq$ 12 (in the case of $U_K$ = $-$ 80 MeV), $K^-$ mesons become unbound, where $\omega_{K^-}\gtrsim m_{\Lambda^\ast}-m_N$ above the $\Lambda^\ast$-resonance region.

From Fig.~\ref{fig3}, the $B/|S|$ steadily increases with $|S|$ in the case in which the SOE is included, while it shows little dependence upon $|S|$ without the SOE.
One finds that $m_{K}-B/|S| > m_{\Lambda({\rm 1116})}-m_N$, where $m_{\Lambda({\rm 1116})}$ is the free mass of the lightest hyperon $\Lambda$(1116). Hence the MKN decays through strong processes such as $K^- N N\rightarrow \Lambda({\rm 1116}) N$, so that it is not stable as a self-bound object. This result qualitatively agrees with that in Gazda et al.\cite{gfgm08}.
\vspace{-0.2cm}

\section{Concluding remarks}
\label{sec:remark}

With regard to creating self-bound objects for the MKN, 
hyperon-mixing effects may be responsible for formation of more strongly bound states. It has been shown in a liquid-drop picture that coexistence of antikaons and hyperons leads to highly dense self-bound objects, which may decay only through weak processes\cite{m08}. There is a controversy about the possible existence of such objects depending on the adopted models and approximations\cite{gfgm09}. A realistic framework including antikaons and hyperons as well as nucleons beyond the local density approximation for baryons is necessary for further investigation.
\vspace{-0.3cm}

\section*{Acknowledgments}
This work is supported in part by the Grant-in-Aid for Scientific Research (No.~20028009).

\end{document}